\documentclass[10pt,twosided,a4paper,draft,onecolumn]{article}
 
\usepackage{microtype}%if unwanted, comment out or use option "draft"
\usepackage[hmargin=1.2in,vmargin=1in]{geometry}
\usepackage{tikz}
\usetikzlibrary{matrix,arrows,decorations.pathmorphing}
\usepackage{times}
\usepackage{epsfig}
\usepackage{cite}
\usepackage{graphicx,color}
\usepackage{amsmath,amsfonts,amssymb,mathrsfs,wasysym,stackrel,amsthm}
\usepackage{algorithmic}
\usepackage{algorithm}
\usepackage{float}
\usepackage{xspace}
\usepackage{multirow,tabularx}
\usepackage{booktabs}
\usepackage{authblk}

\newcommand{\remove}[1]{}

\newcommand{\exprand}[1]{\mathsf{exprand}(#1)}
\newcommand{\abs}[1]{\Bigl\lvert#1\Bigr\rvert}
\newcommand{\Var}[1]{\mathsf{Var}(#1)}

\newcommand{\re}[1]{\mathsf{Real}#1}
\newcommand{\prob}[1]{\mathbb{P}\left[ #1 \right]}
\newcommand{\EXP}[1]{\mathbb{E}\left( #1 \right)}

\newtheorem{theorem}{Theorem}

\newtheorem{corollary}{Corollary}
\newtheorem{Fact}{Fact}

\newcommand{\myunit}{1 cm}
\tikzset{
    node style sp/.style={draw,circle,minimum size=\myunit},
    node style ge/.style={circle,minimum size=\myunit},
    arrow style mul/.style={draw,sloped,midway,fill=white},
    arrow style plus/.style={midway,sloped,fill=white},
}

\title{In-Network Estimation of Frequency Moments\footnote{Pooja
    Vyavahare was supported by DST project
    \textit{SR/S3/EECE/0080/2009} and the work was done in the Bharti
    Centre for Communication at IIT Bombay.  }}

\author[1]{Pooja Vyavahare}
\author[2]{Nutan Limaye}
\author[1]{D. Manjunath}
\affil[1]{Department of Electrical Engineering, IIT Bombay\\
\texttt{vpooja,dmanju@ee.iitb.ac.in}
}
\affil[2]{Department of Computer Science and Engineering, IIT Bombay\\
\texttt{nutan@cse.iitb.ac.in}}

\begin{document}

\maketitle

\markboth{P. Vyavahare et al.}{In-Network Estimation of Frequency Moments}

\begin{abstract}
  We consider the problem of estimating functions of distributed data
  using a distributed algorithm over a network.  The extant literature
  on computing functions in distributed networks such as wired and
  wireless sensor networks and peer-to-peer networks deals with
  computing linear functions of the distributed data when the alphabet
  size of the data values is small, $O(1)$.

  We describe a distributed randomized algorithm to estimate a class of
  \emph{non-linear functions} of the distributed data which is over a
  \emph{large} alphabet. We consider three types of networks:
  point-to-point networks with gossip based communication, random
  planar networks in the connectivity regime and random planar
  networks in the percolating regime both of which use the
  slotted Aloha communication protocol. For each network type, we estimate the scaled $k$-th
  frequency moments, for $k \geq 2$.  For every $k\geq
  2$, we give a distributed randomized algorithm that computes, with
  probability $(1-\delta),$ an $\epsilon$-approximation of the scaled
  $k$-th frequency moment, $F_k/N^k$, using time $O(M^{1-\frac{1}{k-1}} T)$ and 
  $O(M^{1-\frac{1}{k-1}} \log N \log (\delta^{-1})/\epsilon^2 )$ 
  bits of transmission per communication step.
  Here, $N$ is the number of nodes in the network, $T$ is the
  information spreading time and $M=o(N)$ is the alphabet size.
 
\end{abstract}

\section{Introduction}
\label{sec:intro}

We consider the problem of distributed computation of the $k$--th
frequency moment ($k \geq 2$) of data that is distributed over a
network. We assume that there are $N$ nodes in the network and each
node holds a number $x_i$ from a \textit{large} alphabet set
$\mathcal{A} := \{1, \ldots ,M\},$ where $M=o(N).$ If $N_m$ is the
number of times $m \in \mathcal{A}$ appears in the network, then the
$k$--th frequency moment of the data is defined as
$F_k:=\sum\limits_{m=1}^M (N_m)^k.$ The frequency moments are an
important statistic of the input data. $F_0$ is the number of distinct
elements in the data, $F_1$ is the size of the data. $F_2,$ also known
as Gini's index or the `surprise index', is a measure of the
dispersion in the data.  More generally, for $k \geq 2,$ the frequency
moments are an indication of the skewness of the data: $F_k/N^k=1$
indicates a highly skewed data and $F_k/N^k=1/M^{k-1}$ corresponds to
a uniform distribution of the data. Our interest is the case of $k\geq
2$ for which $F_k/N^k$ is in the range $[1/M^{k-1}, 1].$

The estimation of the frequency moments has played a central role in
designing algorithms for database management systems. Many algorithms
for estimating the frequency moments have been considered in the
past. For a detailed survey of the literature we point the reader to
\cite{NelsonThesis11}. In this literature, the main assumption is that
the data is being processed by a single processor. The processor gets
a small snap-shot of the data at any given time and it revisits the
data very few times. The primary focus of the known algorithms such as
those of \cite{Alon96,Ganguly04,Coppersmith04} is to reduce the space
needed to estimate the frequency moments. This is important because
today the data size is massive while
the amount of space available to process them is comparatively small.

In this work, we focus on a setting that is different than that of
\cite{Alon96,Ganguly04,Coppersmith04}. We consider the model in which
the data is distributed among many processors. (The terms processor
and node will be used interchangeably.) We consider the case
in which each processor holds exactly one element of the data and the
processors form a communication network. The rules governing the
communication among the nodes are fixed. Therefore, the algorithm must
work against the given network topology, the given properties of the
network, and the given rules of communication in the network. As the
data is distributed, the parameters of interest are (a) the number of
bits transmitted per node, and (b) the amount of time needed to
compute the estimates of the frequency moments at all the nodes or at
a designated node. Our algorithms optimize both the parameters
simultaneously.

There are several networks where the algorithms like those of
\cite{Alon96,Ganguly04,Coppersmith04} can be used directly. As an
example, consider the case in which each node has a unique identifier
and it can broadcast its data to every other node in the network. In
this case, the task is easy. The algorithm designer can assign the
role of a leader to one of the nodes and assign one slot for each node
to transmit. The nodes then broadcast their data during the assigned
slot. The leader receives the data broadcast by other nodes as a
stream of data. This is identical to the situation of a unique
processor and a massive dataset. The application of the algorithms of \cite{Alon96,Ganguly04,Coppersmith04}
is now obvious. Two network characteristics
complicate matters---(1)~nodes do not have a global identifier, e.g.,
point-to-point networks with gossip based communication (like those considered in
\cite{Boyd05,Mosk-Aoyama06}) and structure-free wireless sensor
networks with slotted Aloha based communication (like that in
\cite{Kamath08}) and, (2)~nodes form a multi hop network (like in
\cite{Boyd05,Mosk-Aoyama06, Kamath08}) possibly with a fraction of the
nodes not being a part of the main connected component (like in
\cite{Iyer11}). In this paper we consider the following three settings
that have these two characteristics and develop randomized algorithms
to obtain estimate of $F_k/N^k.$

\begin{itemize}
\item Point-to-point network with gossip based communication: Here
  every node in the network knows its neighbors and can only
  communicate with them. The network is assumed to form a single
  connected component. At the end of the computation, each node is
  required to know the value of the function. Many recent works have
  considered this setting in which communicating pairs are chosen
  randomly at each time step; time steps are generated by a Poisson
  clock. See for example \cite{Ayaso08,Boyd05,Mosk-Aoyama06}. We will
  refer to these as \emph{gossip networks.} 
\item Random planar radio networks (RPRN) with slotted Aloha
  communication:  The nodes are randomly distributed in the unit
  square. Each node broadcasts its data and all nodes within the
  transmission range of it receive this broadcast data. The efficiency
  of these networks is determined by the spatial reuse factor which is
  inversely proportional to the square of the transmission range. Thus
  we want the transmission range to be as small as possible. However,
  if it is too small, the network will be disconnected and computing a
  global function will be impossible. From \cite{Gupta00}, we know that the
  smallest transmission range for which the network will be a single
  connected component with high probability is $r(N) = \Theta
  \left(\sqrt{\ln N/N}\right).$ This setting of the network (i.e.,
  radius set to $r(N)$) is referred to as the \emph{connectivity
    regime} and we will call them \emph{connected RPRNs.}  This regime
  for function computation has been studied in \cite{Kamath08,
    Giridhar06, Dutta08}. For networks in which the nodes have a
  global identity, like in the models of \cite{Giridhar06}, a trivial
  extension to the algorithms for type sensitive functions (as in
  \cite{Giridhar05}) can be used.

\item Percolating RPRNs with slotted Aloha communication: This is
  similar to the preceding setting except that the transmission range
  is smaller and is chosen to produce a single \emph{giant component}
  in the network rather than a single connected component. In this
  regime the network will have several \emph{smaller} components in
  addition to the giant component. Computation will be performed in
  the giant connected component. We will call this setting
  \emph{percolating RPRNs.} Since this component does not contain a
  constant fraction of the nodes, there is data loss and the
  computation is necessarily approximate. The quality of this
  approximation can be controlled by a suitable choice of the
  transmission range which will be $\Theta \left(1/\sqrt{N} \right).$
  Such a setting has been considered recently in \cite{Iyer11}.
\end{itemize}

In networks without global identifiers, a straightforward randomized
algorithm to compute any function is as follows. Let each node pick a
number independently and randomly from a large enough range (say
$4N^3$). By a union bound, each node will have a unique identifier
with high probability. Now any algorithm which works with node
identifiers will work. Therefore, for point-to-point (random planar)
networks, we will be able to design a randomized algorithm which
transmits $O(N^3 \log M)$ bits per processor. We do not know any obvious
technique to reduce the number of bits transmitted per
processor. However, our goal is to design algorithms that transmit $o(N)$ bits per processor. 

As pointed out by \cite{Giridhar06}, and to the best of our knowledge,
all of the literature on in-network function computation aims to
compute linear functions of the data distributed over the network such as the
sum of the data values or average of the data values. Also, they work
for a small input alphabet, i.e., $M=O(1)$. We break away from both
these restrictions.

\subsection*{Our Contributions:}
\begin{itemize}
\item We give algorithms to estimate scaled frequency moments in the
  three types of networks listed above. To the best of our knowledge,
  this is the first work which estimates a class of non-linear functions in such
  networks.

\item We also get rid of the standard restriction that $M=O(1)$. We
  allow $M \rightarrow \infty$. The only constraint we have on $M$ is
  $M=o(N)$.
\end{itemize}
We achieve this by using two
techniques---(1)~\emph{sketching} which is a standard tool in many
randomized algorithms (e.g., \cite{Muthukrishnan05,Motwani96}), and
(2)~\emph{exponential random variables}, which were first introduced
in distributed computing by \cite{Cohen97} and later used by many
other works including those on gossip based computation (e.g.,
\cite{Mosk-Aoyama06}).

Intuitively, the technique of sketching reduces the problem of size
$M=o(N)$ to that of $M=O(1)$. This alone does not suffice. We observe
that the existing sketching algorithms for computing frequency moments
have some additional properties, which help us compose exponential
random variables with the random maps used for sketching. These two
maps give a small set of random variables. We analyze the properties
of these random variables to finally obtain our results.  The main
theorems in our paper can be stated as follows:

\begin{theorem}
\label{thm:main}
For all constants $\epsilon, \delta \in (0,1)$, there
exist $r_1,r_2 = poly(\epsilon^{-2}, \log \delta^{-1})$ such that there is
 a randomized algorithm that runs in time $O(T)$, uses $O(r_1r_2\log N)$ bits of
transmission per step, and computes an estimate of
$\frac{F_2}{N^2}$, say $f_{2}$, such that $\prob{ |f_2 -
  \frac{F_2}{N^2}| \geq \epsilon} \leq \delta.$ Here, $T$ is equal to
$T_1$ for gossip networks and $T_2$ for connected RPRNs.

\end{theorem}

\begin{theorem}
\label{thm:Fk}
For all $k \geq 3$ and for all constants $\epsilon, \delta \in (0,1)$,
there exist $r_1,r_2 = poly(\epsilon^{-2}, \log \delta^{-1}),$ $B\cdot s_1 = O(M^{1-\frac{1}{k-1}})$
such that there is a randomized algorithm that runs in time $O(Bs_1T),$ uses $O(r_1r_2\log N)$ bits of transmission per step
and computes an estimate of $\frac{F_k}{N^k}$, say ${f}_{k}$, such that\\
$\prob{ |f_k - \frac{F_k}{N^k}| \geq \epsilon} \leq \delta.$ Here, $T$
is equal to $T_1$ for gossip networks and $T_2$ for connected
RPRNs.

\end{theorem}

\section{Preliminaries}
\label{sec:prem}
In this section, we formalize many of the notions described in the
preceding section. We also list a few known definitions and theorems
which we will use in the subsequent sections.
\subsection{The model}
\label{sec:model}

We assume that there are $N$ nodes in the network and the value of $N$
is known to all the nodes. Let $x_u \in \mathcal{A}$ be the data at
node $u.$ Without loss of generality, we assume that $\mathcal{A} =
[1,\ldots, M].$ We further assume $M=o(N)$ and define $x:=(x_1,x_2,
\ldots x_N).$ As we mentioned earlier, we consider three different
types of network models.

Assume that the computation starts at time $0.$ At any time $t > 0,$
each node would have an intermediate function that is determined by a
subset of the nodes in the network. Let $f_i(t)$ denote this function
at node $i$ at time $t.$ 

\begin{itemize}
\item Gossip networks: Here the nodes know their neighbors, but not
  the entire network topology; the nodes do not have global
  identifiers. The communication model is as follows. There is a
  global Poisson clock ticking at rate $N$ per unit time. A random
  communication and a corresponding computation event is scheduled at
  each tick of the Poisson clock. At each clock tick, a node is
  selected uniformly at random from among the $N$ nodes and the node
  performs a communication and a computation operation with randomly
  selected neighbor. The number of bits to be exchanged in each communication and computation operation 
  will be determined in Section~\ref{sec:err-analysis}. 
  The time for the algorithm to complete does not include the time to exchange data.   
  This communication model is called the gossip mechanism. The goal is to compute an 
  estimate of the function value at each node using such gossip
  communication model. This model is fairly well known and is
  described in detail in, among others, \cite{Shah09,Boyd05}.

  Let $S_u(t)$ be the set of nodes that have the data $x_u$ and/or used it to compute 
  their function at time $t,$ i.e.,
  \begin{displaymath}
   S_u(t) := \{v: \mbox{node $v$ has the value $x_u$ and/or used $x_u$ to compute $f_v(t)$}\}
  \end{displaymath}
  In a gossip algorithm at a clock tick at time $t,$ if edge $(u,v)$ is chosen, then the
  data of the nodes that have been used to compute $f_u(t^{-})$ 
  (which is the function value at $u$ immediately before time $t$) would now
  determine $f_v(t)$ and likewise for $f_v(t).$
  If $v \in S_u(t)$, then we say that $v$ has \emph{heard} $u$ before
  time $t.$ The information spreading time is defined as,
  \begin{displaymath}
    T_1 := \inf \{t: \prob{ \cup_{u=1}^N \{ |S_u(t)| \neq N \}} \leq
    \beta_1 \}. 
  \end{displaymath}
  Here the probability is over the randomness of the communication
  algorithm. In other words, $T_1$ is the minimum time required so
  that the event ``every node has heard every other node'' has
  occurred with probability at least $(1-\beta_1).$ As shown in
  \cite{Mosk-Aoyama06} $T_1$ depends on total number of nodes in the
  network and also on how well the network is connected. Specifically, 
  \begin{displaymath}
    T_1 = O \left(N\frac{\log N + \log \beta_1^{-1}}{\Phi(P)} \right),
  \end{displaymath}
  where, $P$ is the adjacency matrix of the graph and $\Phi(P)$ is the
  conductance of $P.$

\item Connected random planar radio networks: In this case nodes are
  deployed randomly in a unit square and a graph is formed by
  constructing edges between all pairs of nodes which are at most
  $r(N)$ distance apart. Transmission of a node $u$ is received by all
  the nodes $v$ whose distance from $u$ is less than $r(N).$ It has
  been shown in \cite{Gupta00,Penrose03} that if $r(N) = \Theta
  \left(\sqrt{\ln N/N}\right),$ then the network is connected with
  high probability. This choice of $r(N)$ corresponds to the
  connectivity regime. The communication algorithm used here is the
  slotted Aloha protocol---at any time step $t$ each node transmits
  with probability $p_N$ and if it transmits, it will transmit the
  current information that it holds. Multiple bits can be transmitted
  in each slot; the exact number will be determined in Section~\ref{sec:err-analysis}.
  The objective is to compute an estimate of the function at all the
  nodes. If $p_N$ is chosen suitably then it can be ensured that the
  transmission from any node will be received by at least one of its
  neighbors with a constant probability, independent of $N.$ See
  \cite{Kamath08} for a more detailed discussion on such information
  spreading algorithms.

  Consider a time slot $t$ in which node $u$ transmits and it is
  received correctly by node $v.$ Node $u$ would be transmitting
  $f_u(t-1).$ Clearly, the data from the nodes that were used to compute
  $f_u(t-1)$ would now determine $f_v(t).$ Let $S_u(t)$ be defined as
  before; the information spreading time, $T_2$ can also be defined
  similarly, i.e., 
  \begin{displaymath}
    T_2 := \min \{t: \prob{ \cup_{u=1}^N \{ |S_u(t)| \neq N \}} \leq
    \beta_2 \}. 
  \end{displaymath}
  where the probability is over the randomness in the slotted Aloha
  protocol. It has been shown in \cite{Kamath08} that if $p_N =
  \Theta(1/ \log N)$ and $r_N = \Theta \left(\sqrt{\ln N/N} \right),$
  then
  \begin{displaymath}
    T_2 = O \left(\sqrt{\frac{N}{\log N}} + \log \beta_2^{-1} \right).
  \end{displaymath}

\item Percolating random planar radio networks: Connected RPRNs have
  transmission range $r(N) = \Theta\left(\sqrt{\log N/N} \right).$
  This means that the average degree of a node is $\Theta(\log N).$
  Higher degree reduces spatial reuse factor, i.e., only
  $\Theta\left(N /\log N \right)$ nodes can transmit
  simultaneously. In \cite{Iyer11} it is shown that choosing $r(N) =
  \Theta(1/\sqrt{N})$ and suitably deleting a small number of nodes
  from a random planar network yields a giant component with all the
  nodes having a constant degree. Also, the number of nodes in this
  giant component will be exponentially larger than the second largest
  component. In fact, $r(N)$ can be chosen to ensure that the giant
  component has at least a specified fraction, say $(1-\alpha)$ (where
  $0 < \alpha <1$), of the nodes. We will perform the computation in
  this giant component. Since the nodes which are not in the giant
  component do not participate in the computation, the computation is
  necessarily approximate. The analysis of this network will follow
  that of connected RPRNs very closely. We will not elaborate on this
  here.
\end{itemize} 

\subsection{Frequency moments}
\label{sec:prem_fk}

Recall that if $N_m$ is the number of times $m$ appears in the network
then the $k$--th frequency moment of $x$ is $F_k:=\sum\limits_{m=1}^M
(N_m)^k.$ A randomized algorithm to estimate $F_2$ is given in
\cite{Alon96}. This works in the situation where there is a single
processor. To design our distributed algorithm, we use the random maps
from their algorithm. To make the description self-contained, we
recall their algorithm:
\begin{algorithm} 
  \caption{Streaming algorithm to compute $F_2$}{\label{alg:f2_stream}} 
  \begin{algorithmic}[1]
    \REQUIRE $x \in \{1,2,\ldots,M\}^N$,  4-wise independent maps
    $\phi_1, \ldots ,\phi_{r_1} : \mathcal{A} \rightarrow \{+1,-1\}$  
    \STATE  $y_u^i \gets \phi_{i}(x_u)$, $1 \leq i \leq r_1$ and $1
    \leq u \leq N$ 
    \STATE $s^i \gets \sum\limits_{u=1}^N y_u^i,$  $1 \leq i \leq
    r_1.$ 
    \STATE  $\hat{F}_2 \gets \frac{\sum\limits_{i=1}^{r_1} (s^i)^2}{r_1}$ 
  \end{algorithmic}
\end{algorithm}

The following theorem characterizes the performance of Algorithm~1.
\begin{theorem}
  \label{thm:ams-F2}
  (\cite[Theorem 2.2]{Alon96} )~~$\EXP{\hat{F}_2} = F_2,$ and
  $\Var{\hat{F}_2} \leq 2F_2^2$ and hence

  \begin{displaymath}
    \prob{(1-\epsilon)F_2 \leq \hat{F}_2 \leq (1+\epsilon)F_2} \geq
    1-\frac{2}{r_1\epsilon^2}
  \end{displaymath}
\end{theorem}

Observe that $s^i = (N_+^i -N_-^i) = (N_+^i - (N-N_+^i)) = (2N_+^i
-N),$ where $N_+^i$ is the number of elements mapped to $+1$ under the
map $\phi_i$ and $N_-^i$ is the number of elements mapped to $-1$
under the map $\phi_i.$ Hence to compute $\hat{F}_2$ the algorithm
requires only the number of elements mapped to $+1.$

\subsection{Exponential random variables}
\label{sec:exp}

Let $X \sim \exprand{a}$ be exponentially distributed random variable
with mean $1/a$. The probability distribution function corresponding
to $X$, denoted as $g_X(x),$ is defined as:
\begin{equation*}
  g_X(x) = \begin{cases}
    ae^{-ax} & \mbox{if } x\geq 0 \\
    0 & \mbox{if } x <0
  \end{cases}
\end{equation*}
\begin{Fact} 
  \label{fact_exp}
  (\cite[Property 1]{Mosk-Aoyama06}) Let $X_i \sim \exprand{a_i}$ be
  independent exponential random variables with mean $1/a_i$ for each
  $i \in \{1, \ldots N\}.$ Let $\tilde{X} =\min_{i=1}^{N}X_i.$ Then
  $\tilde{X}$ is also an exponential random variable with mean
  $\left( \sum\limits_{i=1}^N a_i\right)^{-1}.$
\end{Fact}

\section{Algorithm for Second Frequency Moment}
\label{sec:F2}
\subsection{Algorithm}
\label{sec:f2_algo}

\begin{figure}[h]
\begin{center}
\begin{tikzpicture}[>=latex]
 \tikzstyle{every node} = [circle,fill=gray]
 \node (a) at (2.5,2) {$x_u$};
 \node (b) at (0,0) {$y_u^1$};
 \node (c) at (5,0) {$y_u^{r_1}$};
 \node (d) at (-1.5,-1.5) {$z_u^{1,1}(0)$};
 \node (e) at (1.5,-1.5) {$z_u^{1,r_2}(0)$};
 \node (f) at (3.5,-1.5) {$z_u^{r_1,1}(0)$};
 \node (g) at (6.5,-1.5) {$z_u^{r_1,r_2}(0)$};
 \draw [->] (a) -- (b) node[pos=.5,sloped,above,fill=white] {$\phi_1$};;
 \draw [->]  (a) -- (c)node[pos=.5,sloped,above,fill=white] {$\phi_{r_1}$};;
 \draw [->] (b) -- (d);
 \draw [->] (b) -- (e);
 \draw [->] (c) -- (f);
 \draw [->] (c) -- (g);
 \fill[black] (1,0) circle (0.3ex);
 \fill[black] (2,0) circle (0.3ex);
 \fill[black] (3,0) circle (0.3ex);
 \fill[black] (4,0) circle (0.3ex);
 \fill[black] (-0.5,-1.5) circle (0.3ex);
 \fill[black] (0.5,-1.5) circle (0.3ex);
 \fill[black] (4.5,-1.5) circle (0.3ex);
 \fill[black] (5.5,-1.5) circle (0.3ex);
 \fill[black] (2.5,-1.5) circle (0.3ex);
\end{tikzpicture}
\end{center}
\caption{Mapping to $r_1r_2$ random variables in a node with data $x_u$}
\label{fig:map}
\end{figure}
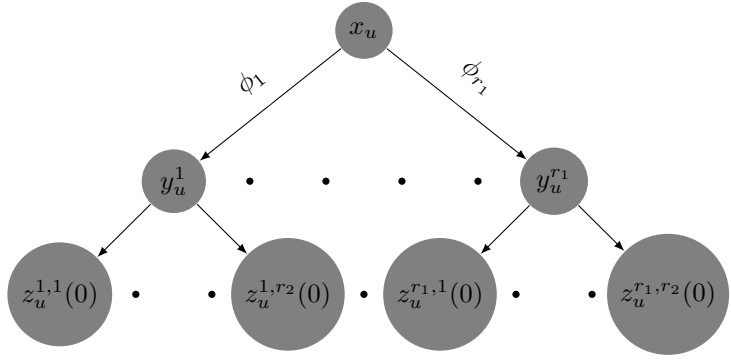

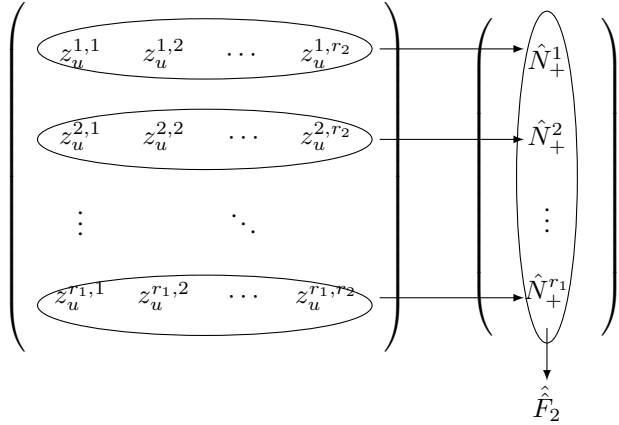
\begin{figure}[h]
\begin{center}
\begin{tikzpicture}[>=latex]
 \matrix (A) [matrix of math nodes,%
             nodes = {node style ge},%
             left delimiter  = (,%
             right delimiter = )] at (0,0)
             {
 z_u^{1,1} & z_u^{1,2} & \cdots & z_u^{1,r_2} \\
 z_u^{2,1} & z_u^{2,2} & \cdots & z_u^{2,r_2} \\
 \vdots & & \ddots \\
 z_u^{r_1,1} & z_u^{r_1,2} & \cdots & z_u^{r_1,r_2} \\
};
%%%
 \matrix (B) [matrix of math nodes,%
             nodes = {node style ge},%
             left delimiter  = (,%
             right delimiter = )] at (4.5*\myunit,0)
             {
  \hat{N}_+^1 \\
  \hat{N}_+^2 \\
  \vdots \\
  \hat{N}_+^{r_1} \\
};
%%%
 \draw  (0,1.7) ellipse (2.2cm and 0.4cm);
 \draw [->] (2.3,1.7) -- (4.2,1.7);
 \draw  (0,0.5) ellipse (2.2cm and 0.4cm);
 \draw [->] (2.3,0.5) -- (4.2,0.5);
 \draw (0,-1.7) ellipse (2.2cm and 0.4cm);
 \draw [->] (2.3,-1.6) -- (4.2,-1.6);
 \draw (4.5,0) ellipse (0.4cm and 2.2cm);
 \draw [->] (4.5,-2) -- (4.5,-2.7);
 \node at (4.5,-3) {$\hat{\hat{F}}_2$};
\end{tikzpicture}
\end{center}
\caption{Estimating $N_+$ and $F_2$}
\label{fig:estimate}
\end{figure}

Our algorithm has three parts. The first part consists of computations
performed per node depending on its own data. In this part, first
every node $u$ maps its data $x_u$ to $r_1$ random numbers $\{y_u^{1}
, \ldots ,y_u^{r_1} \}$ using independent random maps and then each of
the $y_u$'s are mapped to $r_2$ independent random variables. Thus
each node $u$ maps $x_u$ to $r_1r_2$ random numbers as shown in
Figure~\ref{fig:map}. The second part involves exchange of information
across the network to compute a function $\{z_u^{1,1} , \ldots ,
z_u^{r_1,r_2}\}$ of the random numbers generated in the first step. In
the last stage the $z_u$'s are first used to estimate intermediate
estimators $\{\hat{N}_+^1, \ldots, \hat{N}_+^{r_1}\}$ and finally an
estimate of $F_2$ is calculated as shown in Figure~\ref{fig:estimate}.
The exact procedure is explained in Algorithm~\ref{alg:f2}.

%%%%%%%%%%%%%%%%%%%%%%
%%%%%%%%%%%

\begin{algorithm} 
  \caption{Algorithm run by node $u$}{\label{alg:f2}}
  \begin{algorithmic}[1] 
    \REQUIRE $x_u \in \mathcal{A}$, independent maps $\phi_1, \ldots ,
    \phi_{r_1} : \mathcal{A} \rightarrow \{+1,-1\}$ 
    \STATE $y_u^i \gets \phi_{i}(x_u)$, $1 \leq i \leq r_1$ 
    \STATE For each $j \in [1,r_2],$ $z_u^{i,j}$ is chosen randomly
    and independently according to 
    \begin{equation*}
      z_u^{i,j}(0) \gets \begin{cases}
        \exprand{y_u^i} & \mbox{if } y_u^i =1 \\
        \infty & \mbox{if } y_u^i = -1
      \end{cases}
    \end{equation*}
    \STATE Depending on the information spreading algorithm, node $u$
    receives information from node $v$ at time $1 \leq t \leq T.$ On
    receipt of this information it updates as follows:
    \begin{displaymath}
      z_u^{i,j}(t) \gets \min\{z_u^{i,j}(t-1),z_v^{i,j}(t-1)\}
    \end{displaymath}
    \STATE Let $\hat{N}_+^{i} \gets  r_2 \ \left(
      \sum \limits_{j=1}^{r_2} z_u^{i,j}(T) \right)^{-1}.$ 
    \STATE $\hat{\hat{F}}_2 \gets r_1^{-1} \ \sum \limits_{i=1}^{r_1}
    \left(2\hat{N}_+^i -N \right)^2.$   
  \end{algorithmic}
\end{algorithm}

The mapping of the elements of $x$ using random maps $\phi_i$ are
4-wise independent in \cite{Alon96}. However, in our setting we can
use independent random maps because we are not trying to optimize the
number of bits stored per node. Rather, we are trying to optimize the
number of bits transmitted per processor. The random maps can be
thought of as global randomness shared by all the nodes.

\subsection{Error Analysis}
\label{sec:err-analysis}

Let us examine the properties of $\hat{\hat{F}}_2$ obtained in
Algorithm~\ref{alg:f2}. If $N_+^i$ is exactly known then
from Theorem~\ref{thm:ams-F2} the estimate of $F_2$, defined as
$\hat{F}_2$, can be written as:
\begin{displaymath}
  \hat{F}_2 := r_1^{-1} \sum\limits_{i=1}^{r_1} \left( 2N_{+}^i - N \right)^2.
\end{displaymath}
From Theorem~\ref{thm:ams-F2}, we know that 
\begin{equation}
  \label{eq:p1}
  \prob{(1-\epsilon_1)F_2 \leq \hat{F}_2 \leq (1+\epsilon_1)F_2}
  \ \geq \ 1-\frac{2}{r_1\epsilon_1^2} =: p_1.
\end{equation}

However, we do not know, rather cannot know, $N_+^i$ exactly for any
$i$. In our algorithm, $N_+^i$ is a random variable that depends on
the random map $\phi_i$. Steps 3, 4 serve the purpose of estimating
$N_+^i$ for the maps in Step 1, under the assumption that Step 3 has
taken place without any error.  However, recall that in point-to-point
as well as random planar networks, Step 3 itself uses randomness.

\textbf{Error in Step 3 for point-to-point networks:} We say that an
error has occurred in Step 3, if $\exists u \in \{1,2,\ldots,N\}$ such
that $|S_u(T_1)| \neq N$. Therefore, the probability of error is
bounded by $\beta_1.$

\textbf{Error in Step 3 for random planar networks:} We say that an
error has occurred in Step 3, if $|S_0(T_2)| \neq N$. Therefore, the
probability of error is bounded by $\beta_2.$

Assuming no error takes place in Step 3, we now analyze the error in
$\hat{\hat{F_2}}$. To finally bound the overall error, we trivially
combine errors coming from different steps in the algorithm.

Recall that $z_u^{i,j}(T)$ is an exponential random variable.
 Assuming $z_u^{i,j}(T)$ is correct at time $T$, let us
define $Z^i :=\frac{1}{r_2} \sum\limits_{j=1}^{r_2} z_u^{i,j}(T).$
Conditioned on $N_{+}^{i},$ we can use Chernoff bound analysis to show
that for any constant $\epsilon \in (0,1/2),$
\begin{displaymath}
  \prob{ \abs{Z^i - \frac{1}{N_+^i}} > \frac{\epsilon}{N_+^i}}
  \ \leq \ 2 \exp \left(\frac{-\epsilon^2r_2}{3}\right). 
\end{displaymath}
This can be written as
\begin{equation}
  \prob{ (1-\epsilon_2)N_+^i \leq \frac{1}{Z^i} \leq
    (1+\epsilon_2)N_+^i} \ \geq \ p_2 \label{eq:p2} 
\end{equation}
where $\epsilon_2 := 2\epsilon$ and $p_2 := 1-2
\exp\left(-\frac{\epsilon_2^2r_2}{12}\right).$

Writing $\hat{N}_+^i:=1/Z^i,$ i.e., $\hat{N}_+^i$ is the estimate of
$N_+^i,$ and expanding $\hat{\hat{F}}_2,$ we have
\begin{displaymath}
  \hat{\hat{F}}_2 \ =  r_1^{-1} \ \sum\limits_{i=1}^{r_1}(2\hat{N}_+^i -
  N)^2  \ =  \ r_1^{-1} \ \sum\limits_{i=1}^{r_1}\left(4
    \left(\hat{N}_+^i \right)^2 - 4 N\hat{N}_+^i + N^2\right), 
\end{displaymath}

If $(1-\epsilon_2)N_+^i \leq \hat{N}_+^i \leq (1+\epsilon_2)N_+^i,$
then $\hat{\hat{F}}_2$ can be upper bounded as:
\begin{eqnarray*}
  \hat{\hat{F}}_2 & \leq & r_1^{-1} \ \sum \limits_{i=1}^{r_1}
  \left( 4(1+\epsilon_2)^2 \left( \hat{N}_+^i \right)^2 \ - \  
    4N_+^i(1-\epsilon_2)N  + N^2 \right) \\  
  & = & r_1^{-1} \ \sum \limits_{i=1}^{r_1} \left( 4 \left({N}_+^i
    \right)^2 \ - \ 4NN_+^i +N^2 \right) \ + \ r_1^{-1} \ \sum
  \limits_{i=1}^{r_1}\left(4\epsilon_2N_+^i (\epsilon_2N_+^i + 
    2N_+^i +N) \right) \\ 
  & \leq & \hat{F}_2 +  4\epsilon_2N(\epsilon_2N + 2N +N)\\ 
  & \leq & (1+\epsilon_1)F_2 + 4\epsilon_2N(\epsilon_2N + 2N +N) \\
  & \leq & F_2 +N^2((\epsilon_1 + 4\epsilon_2(3+\epsilon_2)) \ = \ F_2
  + N^2 \epsilon  
\end{eqnarray*}
where $\epsilon = \epsilon_1 + 4\epsilon_2(3+\epsilon_2).$ Similarly
we can lower bound $\hat{\hat{F}}_2$ as:
\begin{eqnarray*}
  \hat{\hat{F}}_2 &\geq F_2 - \epsilon N^2.
\end{eqnarray*}
Combining all the three error probabilities i.e., $p_1,$ $(1-\beta),$
$p_2$ corresponding to $\phi$ maps, information spreading algorithm
and exponential random maps respectively, we get, 
\begin{displaymath}
  \prob{\frac{F_2}{N^2} - \epsilon \leq \frac{\hat{\hat{F}}_2}{N^2}
    \leq \frac{F_2}{N^2} + \epsilon} \geq p_1p_2(1-\beta).
\end{displaymath}
Note that $\epsilon$ depends on $\epsilon_1$ and $\epsilon_2$ which
can be chosen arbitrarily small. Also $p_1$ and $p_2$ depend on
$(\epsilon_1,r_1)$ and $(\epsilon_2, r_2)$ respectively. Thus these
can also be made arbitrarily small by suitably choosing $r_1$ and
$r_2.$ $\beta$ can also be made arbitrarily small by suitably choosing
$T.$ 

The space analysis for each node is fairly standard. We include it
here for the sake of completeness. Each node transmits a vector of
length $r_1r_2$. Each entry of this vector is an exponential random
variable. Let us assume that $s$ bits suffice to store the exponential
random variables.\footnote{Then $\infty$ in the algorithm can be
  represented using $s+1$ bits.} When node $u$ receives the vector of
$v,$ it computes the coordinate-wise minimum of the $r_1r_2$-element
vector. Only this minimum is stored at every node and transmitted at
each time the node is activated. This suffices because the ``min''
function is unaffected by the sequence in which the different nodes
are heard and also if a node is heard multiple times.  If $s$ bits
suffice for storing exponential random variables, then each node
transmits $O(r_1r_2s)$ bits. $s$ is determined below by suitably
truncating and quantizing the $z_n^{i,j}(t),$ which are exponential
random variables. 

We will show below that $O(\log N)$ bits of precision suffice to store
$z_n^{i,j}(t).$ This will allow $F_2$ to be estimated within the same
factor of approximation with an additional small error. We thus modify
only Step~2 in Algorithm~\ref{alg:f2} as follows:

\begin{algorithm}
  \caption{Modified step 2 of Algorithm~\ref{alg:f2} with quantized
    random variables}{\label{alg:quant}} 
  \begin{algorithmic}[2a] 
    \STATE For $1 \leq j \leq r_2,$ generate $z_u^{i,j}(0)$ by the
    following rule until $z_u^{i,j}(0) \leq L$ 
    \begin{equation*}
      z_u^{i,j}(0) \gets \begin{cases}
        \exprand{y_u^i} & \mbox{if } y_u^i =1 \\
        L & \mbox{if } y_u^i = -1
      \end{cases}
    \end{equation*}
  \item[] Uniformly quantize $z_n^{i,j}(0)$ using $B$ bits.
  \end{algorithmic}
\end{algorithm}
The other steps of the Algorithm~\ref{alg:f2} remain unchanged.  If
the maximum relative error in estimating $N_+^i$ due to truncation is
$\mu$ and $L$ and $B$ are both chosen as $\Theta(\log N),$ then the
estimate of $F_2$ is, following the analysis of \cite{Ayaso08},
\begin{displaymath}
  \prob{\frac{F_2}{N^2} - \epsilon \leq \frac{\hat{\hat{F}}_2}{N^2}
    \leq \frac{F_2}{N^2} + \epsilon} \geq 1-\delta.
\end{displaymath}
Here, $\epsilon = \epsilon_1 + 8\mu(3+2\mu)$ and $\delta =
e^{\frac{-\mu^2r_2}{6}} + \frac{2}{r_1 \epsilon_1^2}
(1-e^{\frac{-\mu^2r_2}{6}} ).$ This means that in gossip networks at each step a node will transmit $r_1r_2 \Theta(\log N)$ bits. 
Further this also tells us that the each slot
of the slotted Aloha protocol should be $r_1 r_2 \Theta(\log N)$ bit
periods.

We have thus proved Theorem~\ref{thm:main}. 

\subsection*{Percolating RPRN}

Let us now consider the computation of the estimate of $F_2$ in
percolating RPRN except that a fixed fraction of the data is missing,
i.e., $N_{\alpha}:=(1-\alpha) N$ of the nodes have participated in the
computation of $F_2.$ Let $F_{2,\alpha}$ be the second frequency
moment calculated from $N_{\alpha}$ nodes.  Let $m_i$ be arranged in
the descending order as $m_{i_1} \geq m_{i_2}\ldots \geq m_{i_M}$.  It
is easy to see that the difference between $F_{2,\alpha}$ and $F_2$
will be maximized when the nodes that are removed had value
$i_1$. Therefore,
\begin{eqnarray*}
  F_{2,\alpha} \leq (m_{i_1} -\alpha N)^2 + \sum\limits_{j=2}^M
  m_{i_j}^2 \leq \sum\limits_{j=1}^M m_{i_j}^2 -2\alpha N m_{i_1}
  +\alpha^2 N^2.
\end{eqnarray*}
If $m_{i_1} \geq \alpha N, $
\begin{eqnarray}
  F_{2,\alpha}  \leq F_2 -2\alpha ^2 N^2 + \alpha^2 N^2 \leq F_2
  -\alpha ^2 N^2.  \label{eq:f2a} 
\end{eqnarray}
If $m_{i_1} < \alpha N$, then similar calculations can be performed to
get the same bound. Let $\hat{\hat{F}}_{2,\alpha}$ be the output of
Algorithm~2 applied on $N_{\alpha}$ nodes. Then by
Theorem~\ref{thm:main} we know,
\begin{equation}
  \prob{|\hat{\hat{F}}_{2,\alpha} - F_{2,\alpha}| \geq
    (1-\alpha)^2N^2\epsilon}  \leq  \delta \label{eq:f2b} 
\end{equation}  
Using equations \ref{eq:f2a} and \ref{eq:f2b}, we get,
\begin{equation}
  \prob{|\hat{\hat{F}}_{2,\alpha} - F_2| \geq N^2\alpha^2(1-\epsilon)}
  \leq \delta \label{eq:f2percolation} 
\end{equation}
Equation \ref{eq:f2percolation} is applicable to the networks which
operate in percolation regime where a constant fraction of the nodes
do not take part in function computation. Therefore, we have the
following corollary,
\begin{corollary}
  \label{corollary:percolation}
  For all constants $\epsilon,\delta \in (0,1)$, there exist $r_1,r_2
  = poly(\epsilon^{-2}, \log \delta^{-1})$ so that there is a randomized algorithm that runs in time $O(T),$ uses $O(r_1r_2 \log N)$ bits of transmission per step and computes an estimate of
  $\frac{F_2}{N^2}$, say $\frac{\hat{\hat{F}}_{2,\alpha}}{N^2}$, as 
  $\prob{ |\hat{\hat{F}}_{2,\alpha} - F_2| \geq
    N^2\alpha^2(1-\epsilon)} \leq \delta$. Here $T$ is the time
  needed by the spread algorithm and $\alpha$ is the fraction of the
  nodes which are not in the giant
  component.
\end{corollary}

\subsubsection{Second frequency moment using bottom-$r_2$ sketch}
\label{sec:bottom_sketch}

In Algorithm~\ref{alg:f2}, for each node $u$ and $1 \leq i \leq r_1$,
$y_u^i$ is mapped to $r_2$ independent random variables. Let $V_u^i :=
(z_u^{i,1}(T), \ldots ,z_u^{i,r_2}(T)),$ $1 \leq i \leq r_1$ denote
the vector computed by the node $u$ after time $T.$ Note that each
$V_u^i$ is an $r_2$ sized vector each element of which is the minimum
of $N$ independent exponential random variables. $V_u^i$ is also known
as $r_2-$mins sketch in the literature \cite{Cohen07}. Recall from
Section~\ref{sec:err-analysis} that each $z_u^{i,j}(T)$ is an
exponential random variable with mean $1/N_+^i$ and thus is used to
estimate $N_+^i.$ Recall, $r_2 \in O(1)$, therefore it is
asymptotically very small. However, in practice, reducing it to a
small constant will help in reducing the amount of randomness used by
the algorithm, and may help in bringing down the number of bits
transmitted per node.

We observe that $r_2$ can be reduced to $1$. This certainly helps in
reducing the number of random maps used per node. However, because of
the manner in which the final estimate is computed, we do not see any
way of saving the number of bits transmitted per node. We use
\emph{bottom-$r_2$ sketch} as defined in \cite{Cohen07}. For each $1
\leq i \leq r_1$, we map $y_u^i$ to a single exponential random
variable $z_u^i$. For a fixed $i$, arrange $z_u^i$'s in a
non-decreasing order, say $z_{l_1}^i \leq z_{l_2}^i \ldots \leq
z_{l_N}^i$. It is shown in \cite{Cohen07} that using $r_2$ smallest
values, a good estimate can be computed.  Therefore, in our case it
will suffice if each node knows these $r_2$ minimum values.  Each node
can do this by keeping track of the $r_2$ smallest values seen so far
for each $i$. This can be done by the following book keeping: Node $u$
holds a vector $V_u^i := (z_u^{i} , \infty, \ldots \infty),$ for each
$1 \leq i \leq r_1$. Node $u$ communicates the vector $V_u^i$ to node
$v$ and updates its vector by appropriately inserting the values from
$V_v^i$ to get the first $r_2$ minimum values available in the network
for each $i.$ At the end, for each $u$ we have $V_u^i(1) \leq V_u^i(2)
\ldots \leq V_u^i(r_2)$ representing the $r_2$ lowest values of the
network. To summarize, each node can estimate $N_+^i$ by generating
only one exponential random variable instead of $r_2$ independent
random variables. However, to compute the bottom-$r_2$ sketch for
estimating $F_2$, each node transfers $r_1r_2$ numbers, i.e.,
$O(r_1r_2\log N)$ bits of transmission per processor.

%%%%%%%%%%%%%%
\section{Algorithm for Higher Frequency Moments}
\label{sec:Fk}

In this section we present an algorithm to compute frequency moments
$F_k,$ for all $k \geq 3.$ In the data streaming literature, many
algorithms are known for computing $F_k.$ (See for example
\cite{Alon96,Coppersmith04,Ganguly04}.) In \cite{Alon96}, sampling is
used for estimating $F_k$ for $k \geq 3$. For the special case of
$k=2$, they give a sketching algorithm as shown in
Algorithm~\ref{alg:f2_stream}. On the other hand,
\cite{Coppersmith04,Ganguly04} use sketching algorithms for estimating
$F_k$. The map $\phi$ in Algorithm~\ref{alg:f2_stream} can be thought
of as a map from the input alphabet to the square roots of unity. A
possible generalization of this for $k \geq 3$ is a map from the input
alphabet to $k$-th roots of unity. In \cite{Ganguly04} it was proved
that maps from the input alphabet to $k$--th roots of unity can be
used for estimating $F_k.$ In order to estimate $F_k$ in our setting,
we use a combination of random maps to $k$--th roots of unity and
exponential random variables. Our primary observation is that Fact
\ref{fact_exp} helps us compose exponential random variables with the
maps to $k$-th roots. This composed map in turn helps in estimating
$F_k$ for $k \geq 3$ in all the three models of distributed networks.
In order to explain the central idea used in estimating $F_k,$ $k \geq
3$, we give a simplified version of our original algorithm:

\begin{algorithm} 
  \caption{Algorithm for higher frequency moments run by node
    $u$}{\label{alg:fk}} 
  \begin{algorithmic}[1] 
    \REQUIRE $x_u \in \mathcal{A}$
    \hspace*{0.5cm}$\phi_1, \ldots \phi_{r_1}: \mathcal{A} \rightarrow
    \{\alpha_1 +i\beta_1, \ldots ,\alpha_k+i\beta_k\},$ where
    $\alpha_l +i\beta_l = e^{2 \pi i l/k}.$  
    \STATE  $y_{u}^p \gets \phi_p(x_u),$ $1 \leq p \leq r_1$ 
    \STATE If $ y_{u}^p = \alpha +i\beta$ then for $1 \leq q \leq r_2$ 
    \begin{align*}
      z_{\alpha,u}^{p,q}(0) \gets \exprand{\alpha+1} \\
      z_{\beta,u}^{p,q}(0) \gets \exprand{\beta+1} 
    \end{align*}
    \STATE Depending on the information spreading algorithm, node $u$
    receives information from  node $v$ at time step $1 \leq t \leq T.$
    On receipt of this information it updates as follows:  
    \begin{align*}
      z_{\alpha,u}^{p,q}(t) \gets
      \min\{z_{\alpha,u}^{p,q}(t-1),z_{\alpha,v}^{p,q}(t-1)\} \\ 
      z_{\beta,u}^{p,q}(t) \gets
      \min\{z_{\beta,u}^{p,q}(t-1),z_{\beta,v}^{p,q}(t-1)\}  
    \end{align*}
    \STATE $Y^p \gets \re{ \left\{
        \left(r_2 \left(\sum\limits_{q=1}^{r_2}
            Z_{\alpha,u}^{p,q}(T)\right)^{-1}
          +i r_2 \left(\sum\limits_{q=1}^{r_2} Z_{\beta,u}^{p,q}(T)\right)^{-1}
        \right)^k \right\}}$ 
    \STATE $\hat{F}_k \gets  r_1^{-1} \sum\limits_{p=1}^{r_1}Y^p$
  \end{algorithmic}
\end{algorithm}

Note the similarity between Algorithm~\ref{alg:fk} and
Algorithm~\ref{alg:f2}. The above algorithm is overly simplified.  It
was observed by \cite{Ganguly04} that sum of $y_u^i$'s when raised to
power $k$ has expectation equal to $F_k$, however its variance is very
large. This problem was resolved by using a bucketing strategy. For
each node $u$, $x_u$ is mapped to one of $\{1,2,\ldots, B\}$ buckets
using $s_1$ different maps: $\chi_1, \chi_2,\ldots,\chi_{s_1}:
\mathcal{A} \rightarrow \{1,2,\ldots, B\}$. (In our setting we can use
independent random maps because we are not trying to optimize the
amount of bits stored per node. We only try to optimize the number of
bits transmitted per node. The random maps can be thought of as global
randomness shared by all the nodes.) It was proved that $B\cdot s_1 =
O(M^{1-\frac{1}{k-1}})$ \cite{Ganguly04}.

The error analysis of the algorithm can be done in the same way as
done for $F_2$ in Section \ref{sec:err-analysis}. It can be shown that
\begin{displaymath}
  \prob{\frac{F_k}{N^k} - \epsilon \leq \frac{\hat{F}_k}{N^k} \leq 
  \frac{F_k}{N^k} + \epsilon} \geq p,
\end{displaymath}
where $\epsilon$ is a function of $k,M$ and errors due $\phi,\chi$ and
exponential random variables.

Similar to the result of $F_2$ in Section \ref{sec:err-analysis}, $p$
is a function of $r_1,r_2$ and errors due to $\phi,\chi$ and
exponential random variables. As seen earlier, $\epsilon$ can be made
arbitrarily small by controlling the errors due to the random maps and
similarly $(1-p)$ can be made arbitrarily small by choosing $r_1,r_2$
appropriately. Proceeding on the lines of proof of Theorem~\ref{thm:main} we get Theorem~\ref{thm:Fk} from here.

\section{Discussion}
\label{sec:dis}
\begin{itemize}

\item In this paper we have considered one-shot computation of $F_k.$
  For random planar networks, it is also of interest to develop
  algorithms to compute $F_k$ for the sequence of $x,$ the data
  vector. In this case the computation of $F_k$ for the different
  elements of the sequence will be pipelined. The techniques of
  \cite{Kamath08,Iyer11} easily extend to this case.

\item Sketching is a commonly used technique in dealing with massive
  data sets.  It involves mapping the given data from a large alphabet
  into a relatively smaller alphabet preserving the relevant
  properties. Let $f: \{1,2,\ldots, M\}^n \rightarrow
  \{1,2,\ldots,M\}$ be a function and let $\phi: \{1,2,\ldots, M\}
  \rightarrow \{1,2,\ldots,k\}$ denote a map used by a sketching
  algorithm to compute $f$. (Note, $M>>k$).  For $1 \leq i \leq k$,
  let $N^{\phi}_i$ denote the number of elements mapped to $i$ under
  the mapping $\phi$.  Suppose for every input $x = (x_1,
  x_2,\ldots,x_n) \in \{1,2,\ldots,M\}^n$, $f(x)$ can be estimated
  using $N^{\phi(x)}_1, N^{\phi(x)}_2, \ldots, N^{\phi(x)}_k,$ then we
  call $f$ to be \emph{sketch type sensitive.}  That is if $f$ is
  sketch-type sensitive, $f$ essentially depends on a type-vector,
  i.e., a vector of length $k$, with each entry $1\leq i \leq k$ in
  the vector corresponding to the number of elements of the original
  alphabet mapped to $i$.  $F_2$ is one such function. As noted in
  Section \ref{sec:F2}, $F_2= (2N_+-N)^2$.  In fact $\forall k \geq
  2$, $F_k$ is sketch type sensitive. We believe that our techniques
  can be used for estimating any sketch type sensitive function.

\item We compute the estimate of the scaled version of $F_k$, i.e.,
  $F_k/N^k$. It will be interesting to estimate $F_k$ itself. Also, in
  our algorithm, we assume that all nodes know sketching functions
  $\phi_i$'s.  An algorithm that does not assume such shared
  randomness will be an improvement over our algorithm.

\item This work shows that the techniques developed for space
  efficient algorithms to compute functions of streaming data can be
  used to reduce the communication in distributed computing of
  functions of distributed data. This connection needs to be explored
  further.
\end{itemize}

\bibliography{function-computation}

\begin{thebibliography}{10}

\bibitem{Alon96}
N.~Alon, Y.~Matias, and M.~Szegedy.
\newblock The space complexity of approximating the frequency moments.
\newblock In {\em Proc. of the 28th Annual ACM Symposium on the Theory of
  Computing STOC}, pages 20--29, 1996.

\bibitem{Ayaso08}
O.~Ayaso, D.~Shah, and M.~A. Dahleh.
\newblock Counting bits for distributed function computation.
\newblock In {\em Proc. of ISIT}, pages 652--656, 2008.

\bibitem{Boyd05}
S.~Boyd, A.~Ghosh, B.~Prabhakar, and D.~Shah.
\newblock Gossip algorithms: Design, analysis and applications.
\newblock In {\em Proc. of IEEE INFOCOM}, pages 1653--1664, Miami, USA, 2005.

\bibitem{Cohen97}
E.~Cohen.
\newblock Size-estimation framework with applications to transitive closure and
  reachability.
\newblock {\em Journal of Computer and System Sciences}, 55(3):441--453,
  December 1997.

\bibitem{Cohen07}
E.~Cohen and H.~Kaplan.
\newblock Summarizing data using bottom-k sketches.
\newblock In {\em Proc. of Principles of Distributed Computing}, pages
  225--234, August 2007.

\bibitem{Coppersmith04}
D.~Coppersmith and R.~Kumar.
\newblock An improved data stream algorithm for frequency moments.
\newblock In {\em Proc. of 15th annual ACM-SIAM symposium on Discrete
  algorithms}, pages 151--156, 2004.

\bibitem{Dutta08}
C.~Dutta, Y.~Kanoria, D.~Manjunath, and J.~Radhakrishnan.
\newblock A tight lower bound for parity in noisy communication networks.
\newblock In {\em Proceedings Nineteenth Annual ACM-SIAM Symposium on Discrete
  Algorithms (SODA)}, pages 1056--1065, San Francisco, CA USA, 2008.

\bibitem{Ganguly04}
S.~Ganguly.
\newblock Estimating frequency moments of data streams using random linear
  combinations.
\newblock In {\em APPROX-RANDOM}, pages 369--380, 2004.

\bibitem{Giridhar05}
A.~Giridhar and P.~R. Kumar.
\newblock Computing and communicating functions over sensor networks.
\newblock {\em IEEE Journal on Selected Areas in Communications},
  23(4):755--764, April 2005.

\bibitem{Giridhar06}
A.~Giridhar and P.~R. Kumar.
\newblock Toward a theory of in-network computation in wireless sensor
  networks.
\newblock {\em IEEE Communications Magazine}, 44(4):98--107, April 2006.

\bibitem{Gupta00}
P.~Gupta and P.~R. Kumar.
\newblock The capacity of wireless networks.
\newblock {\em IEEE Transactions on Information Theory}, 46(2):388--404, March
  2000.

\bibitem{Kamath08}
S.~Kamath and D.~Manjunath.
\newblock On distributed function computation in structure-free random
  networks.
\newblock In {\em Proc. of IEEE ISIT}, Toronto, Canada, July 2008.

\bibitem{Mosk-Aoyama06}
D.~Mosk-Aoyama and D.~Shah.
\newblock Computing separable functions via gossip.
\newblock In {\em Proc. of 22nd ACM Symposium on Principles of Distributed
  Computing (PODC)}, pages 113--122, 2006.

\bibitem{Motwani96}
R.~Motwani and P.~Raghavan.
\newblock Randomized algorithms.
\newblock {\em ACM Comput. Surv.}, 28(1):33--37, March 1996.

\bibitem{Muthukrishnan05}
S.~Muthukrishnan.
\newblock {\em Data streams: Algorithms and applications}.
\newblock Volume 1, number 2 of Foundations and Trends in Theoretical Computer
  Science. now publishers, 2005.

\bibitem{NelsonThesis11}
J.O. Nelson.
\newblock {\em Sketching and Streaming High-dimensional Vectors}.
\newblock Massachusetts Institute of Technology, Department of Electrical
  Engineering and Computer Science, 2011.

\bibitem{Penrose03}
M.~Penrose.
\newblock {\em Random Geometric Graphs}.
\newblock Oxford University Press, 2003.

\bibitem{Iyer11}
D.~Manjunath S.~K.~Iyer and R.~Sundaresan.
\newblock In-network computation in random wireless networks at constant
  refresh rates with lower energy costs: A {PAC} approach.
\newblock {\em IEEE Transacations on Mobile Computing}, 10(1):146--155, January
  2011.

\bibitem{Shah09}
D.~Shah.
\newblock Gossip algorithms.
\newblock {\em Foundations and Trends in Networking}, 3(1):1--125, 2009.

\end{thebibliography}
\bibliographystyle{plain}
\end{document}